\documentclass[preprint, 5p, 11pt, twocolumn, authoryear]{elsarticle}
\setcitestyle{maxbibnames=1}
\usepackage{amssymb}
\usepackage{lipsum}
\usepackage{hyperref}
\usepackage{aas_macros}
\usepackage{longtable}
\usepackage{booktabs}
\usepackage{float}
\usepackage[latin1]{inputenc}
\usepackage{amsthm}
\usepackage{amsmath}
\usepackage{bm}
\usepackage[normalem]{ulem}
\usepackage{makecell}
\usepackage{array}  
\usepackage{threeparttable}
\usepackage{subcaption} 
\usepackage{booktabs}   
\usepackage{threeparttable} 
\usepackage{siunitx}    
\newcolumntype{@}{>{\global\let\currentrowstyle\relax}}
\newcolumntype{^}{>{\currentrowstyle}}
\newcommand{\rowstyle}[1]{\gdef\currentrowstyle{#1}#1\ignorespaces}

\begin{document}

\journal{High Energy Astrophysics}

\begin{frontmatter}

\title{Detection of new very-high-energy  sources  outside the galactic plane in the Fermi-LAT data}
\author[1,2,3,4]{M.\,S. Pshirkov}
\ead{pshirkov@sai.msu.ru}
\author[4]{A.\,S. Kovankin}
\ead{kovankin.as23@physics.msu.ru}
\affiliation[1]{organization={Sternberg Astronomical Institute, Moscow M.V. Lomonosov State University},
            addressline={Universitetskiy Prospekt 13}, 
            city={Moscow},
            postcode={119991}, 
            country={Russia}}

\affiliation[2]{organization={Lebedev Physical Institute, Astro Space Center, Pushchino Radio Astronomy Observatory},
            addressline={Radioteleskopnaya 1a}, 
            city={Pushchino, Moscow reg.},
            postcode={142290}, 
            country={Russia}} 
        
\affiliation[3]{organization={The Institute for Nuclear Research of the Russian Academy of Sciences},
            addressline={B-312, Prospekt 60-letiya Oktyabrya, 7a}, 
            city={Moscow},
            postcode={117312}, 
            country={Russia}}  
            
\affiliation[4]{organization={Faculty of Physics, Moscow M.V. Lomonosov State University},
            addressline={Leninskie gory 1}, 
            city={Moscow},
            postcode={119991}, 
            country={Russia}}

\begin{abstract}
We present a search for spatio-temporal clusters in 16 years of Fermi-LAT very-high-energy (VHE; $E>100$~GeV) data using the DBSCAN algorithm, focusing on high Galactic latitude ($|b|>10^{\circ}$) clusters with $\geq5$  events and transient doublets (two events within $\leq 3$ days). Of 107 detected clusters, two correspond to previously unidentified VHE sources: weak BL Lacertae objects 4FGL J0039.1-2219 and 4FGL J0212.2-0219, promising targets for next-generation VHE observatories.

Due to the low VHE photon background, even  doublets with a duration of several days exhibited high statistical significance. While most of the 114 detected doublets originated from bright TeV emitters (e.g., Mrk 421, Mrk 501), we identified six VHE flares lacking TeVCat associations. Five of these flares correlate with sources from the Third Catalog of Fermi-LAT High-Energy Sources (3FHL), while one 'orphan' flare lacks a high-energy (HE; $E > 10$~GeV) source counterpart. Some of these flares reached extreme luminosities of $\mathcal{O}(10^{47}~\mathrm{erg~s^{-1}})$.

No consistent temporal correlation emerged between HE and VHE activity: HE flares preceded, coincided with, or followed VHE emission across sources, with some showing no HE counterpart. Remarkably, 3FHL J0308.4+0408 (NGC 1218) is  a Seyfert Type I galaxy,  while no object of this class was known as a VHE emitter before. The 'orphan' flare without any known HE source in the vicinity may originate from NGC 5549, a low-luminosity LINER galaxy. Both sources expand the limited sample of non-blazar AGN detected at VHE energies. The fact that some weak sources  with non-aligned jets and, sometimes, even without any traces of HE activity, could demonstrate very short and powerful VHE flares cannot be easily accounted in many current AGN models and calls for  their further  development.

\end{abstract}
\end{frontmatter}

\section{Introduction}
\label{introduction}

Despite significant observational progress in recent decades, the very-high-energy (VHE; $E>100$~GeV) sky remains understudied compared to other wavelengths and energy ranges. Imaging atmospheric Cherenkov telescopes (IACTs) -- the most sensitive instruments at these energies -- unfortunately have very small fields of view (FoV), which makes conducting deep all-sky surveys extremely challenging.
Observations by the Fermi Large Area Telescope \citep[Fermi-LAT;][]{Atwood2009} could be used to partially alleviate the problem: although its effective area is relatively modest ($\mathcal{O}(1~\mathrm{m^2})$),  Fermi-LAT achieves a substantial total exposure -- approximately $\mathcal{O}(10^{12}~\mathrm{cm^2~s})$ -- owing to its exceptionally wide field of view (2.4 sr) and uninterrupted operation over 16 years. For context,  it is equivalent to a few hours of HESS observations, given that at these energies,  HESS  has  an effective area of $\mathcal{O}(10^4~\mathrm{m^2})$ \citep{Aharonian_2006}. Furthermore, Fermi-LAT's high  cadence of observations provides a unique capability to detect short-lived VHE flares, including rare 'orphan' flares that lack counterparts at other electromagnetic wavelengths. This combination of long time span of observation, all-sky coverage, and high temporal resolution makes Fermi-LAT a critical tool for studying transient and variable VHE phenomena. 

In this paper, we present a search for VHE $\gamma$-ray clusters and transients at Galactic latitudes $|b|>10^\circ$  using 16 years of Fermi-LAT data, focusing on out-of-plane sources likely of extragalactic origin. Many of these sources belong to the high-synchrotron-peak (HSP) blazar class \citep{Tavecchio1998}, where particles accelerated to extreme energies in active galactic nucleus (AGN) jets produce a spectral energy distribution (SED) dominated by synchrotron emission peaking in X-rays and inverse-Compton scattering at VHE energies. These extreme astrophysical accelerators are not only critical for AGN studies \citep{Costamante2001} but also serve as probes for the intergalactic magnetic fields \citep{Tavecchio2011, Khalikov2021}, the extragalactic background light \citep{HESS2017}, and physics beyond the Standard Model \citep{Abdalla2021}.

Fermi-LAT's continuous monitoring capabilities have been leveraged for VHE source detection since the mission's early phases: an analysis of just 1.5 years of data yielded a significant detection of VHE emission from the galaxy IC 310 \citep{Neronov2010} and 5 new Galactic sources at $|b|<10^\circ$ \citep{neronov2010galacticsourcese100gev}. In  5 years of observations, Fermi-LAT detected thirteen high-redshift ($z > 0.5$) sources \citep{Neronov_2015} and 2 VHE photons from the distant ($z = 1.1$) blazar PKS 0426-380 \citep{Tanaka_2013}. Subsequent studies expanded searches to lower energies (tens of GeV or below) using clustering algorithms such as the minimal spanning tree \citep[MST;][]{Campana_MST} and density-based spatial clustering \citep[DBSCAN;][]{Tramacere_2013}. These efforts, combined with accumulating data, produced the 12-year MST catalog of 1888 $\gamma$-ray source candidates above 10 GeV at $|b|>20^{\circ}$ \citep{Campama_mst_catalog_old, Campana_mst_catalog}, identified eight new candidates of supernova remnants in the Large Magellanic Cloud \citep{tramacere2025searchgammarayemissionsnrs}, and revealed 18  sources without counterparts in the Fermi-LAT sources catalog \citep{Hedayati}.

Despite these advances, the wealth of Fermi-LAT VHE data remains underexplored. Here, we apply the DBSCAN algorithm to systematically search for transients and clustered $\gamma$-ray signals, prioritizing candidates unassociated with known VHE sources. This approach aims to uncover new phenomena and refine existing constraints on extreme astrophysical environments.

The paper is organized as follows. In Section \ref{sec:data} we describe our selection of data and the algorithms that   we used in our paper. In Section \ref{sec:results} we present our results and we conclude   in Section \ref{sec:conclusions}.

\section{Data selection and algorithm of cluster detection}
\label{sec:data}
\subsection{Fermi-LAT dataset}
We analyzed 16 years of Fermi-LAT data spanning 4 August 2008 to 9 August 2024, selecting SOURCE-class events with energies $>100$~GeV. The initial dataset contained 47,764 photons. To mitigate contamination from the Galactic diffuse gamma-ray background, we excluded events within $|b|<10^{\circ}$ of the Galactic plane, as this region exhibits significantly higher background flux that reduces sensitivity to faint sources. After applying this latitude cut, the final sample comprised 21,686 VHE photons.

\subsection{Description of the  \emph{DBSCAN} method}
\label{subsec:dbscan}

DBSCAN is a density-based clustering method that uses the local density of data points to find clusters in data sets taking into account the background noise. 
The algorithm groups data points, utilizing two key parameters: $\varepsilon$ (the maximum distance defining the neighborhood of a point) and $K$ (the minimum number of points required to form a cluster core). For each point, the algorithm counts the number of neighbors within its $\varepsilon$-neighborhood. If a point satisfies the core condition (i.e. the number of neighbors is greater than or equal to $K$, including itself), it is designated as a core point and initiates a new cluster including the neighboring points. Clusters are formed by connecting core points and their reachable neighbors. Points within the $\varepsilon$-neighborhood of a core point but with fewer than $K$ neighbors themselves are called border points. They are part of a cluster but cannot expand it. Points unassociated with any cluster are
classified as noise.  
We implemented DBSCAN via the scikit-learn Python package\footnote{\url{https://scikit-learn.org/stable/modules/generated/sklearn.cluster.DBSCAN}}.

As a first approximation we choose $K$ and $\varepsilon$ as follows: 
\begin{enumerate}
    \item Set $\varepsilon = 0.3^{\circ}$ which is close to the instrumental point spread function (PSF) at a given threshold energy (for $E>100$ GeV, $\theta_{\mathrm{PSF}} = 0.1^{\circ}$) 
    \item  To minimize false positives, we derived $K$ using Poisson statistics. For the average background level $\lambda$: 
    \begin{equation}
     \lambda = N/ \Omega_{|b|>10^\circ},
    \end{equation}
    where $N = 21686$ is the  total number of analyzed photons, $\Omega_{|b|>10^\circ} = 4 \pi(1-\sin{10^\circ})$ is the  solid angle of the survey, the probability that $k$ photons fall within  a solid angle $\Omega_{\varepsilon}$ is described by the Poisson distribution:
    \begin{equation}
     P_{\mathrm{loc}}(k) = \dfrac{(\lambda \Omega_{\varepsilon})^k \exp{(-\lambda \Omega_{\varepsilon})}}{k!}
    \end{equation}

    Finally, we used the Bonferroni correction, accounting  for a large number of trials, so the global false-alarm probability became:
    \begin{equation}
        P_{\mathrm{glob}}(k) = \dfrac{\Omega_{|b|>10^\circ}}{\Omega_\varepsilon} P_{\mathrm{loc}}(k) 
        \label{eq:pglob}
    \end{equation}
    where $\Omega_{|b|>10^\circ}/\Omega_\varepsilon$ was the number of analyzed fields. For the given $\lambda$ and $\varepsilon$  $P_{\mathrm{glob}}(k)$ depends on $k$ as follows: 
\begin{table}[h!]
\centering
\begin{tabular}{|c|c|c|c|c|} 
\hline
$k$ & 4&5&6&7  \\ 
 \hline
 $P_{\mathrm{glob}}$ &4.39&0.158&0.00474 &0.00012 \\ 
\hline
\end{tabular}
\end{table} \\
Based on this, we adopted $K=5$ ($P_{\mathrm{glob}}\approx0.16$) for an initial clustering. Final candidate clusters were re-evaluated individually for their statistical significance.

\item\label{it:newpglob} The  average background level was calculated for each cluster using the following procedure: we considered a  $10^\circ \times  10^\circ$ square centered on the center of the cluster $(l_0,~b_0)$ (if $|b_0| < 20^\circ$, we take $b \in [10^\circ,~ 20^\circ]$ or $b \in [-20^\circ,~ -10^\circ]$, if $b_{0}<0^\circ$), after that we  excluded all photons within $0.5^\circ$ of any 3FHL source in this square and obtain a new number of background photons $N'$ and an average background level $\lambda'$.

The refined background density $\lambda'$ and the cluster solid angle $\Omega_{\mathrm{cls}} = \pi \rho^2_{\mathrm{max}}$  (where $\rho_{\mathrm{max}}$ is the cluster's maximum angular radius) were substituted into Eq.~\ref{eq:pglob} to calculate updated $P_{\mathrm{glob}}$ values.

\end{enumerate} 

\subsection{Extended cluster analysis}
Clusters exhibiting significant spatial extension may originate from non-pointlike sources. To quantify this, we compare their photon distributions to the distribution of photons from the  Crab Nebula and pulsar (close to  a point source for our purposes at very high energies, with a radius of $\sim1.5'$ \citep{HESS_Crab2024}). The Crab reference cluster contains 376 photons (210 front-converted, 166 back-converted), providing an empirical template for the Fermi-LAT point-spread function (PSF) at $E>100$~GeV.

\begin{enumerate}
 
\item \textbf{Photon Properties and Empirical PSF:}
For each photon in a cluster, we calculate its angular distance $r$ to the candidate source position (or cluster centroid). The probability $p_i(r)$ of observing a photon of conversion type $i$ (front/back) at a distance $\geq r_0$ is:
$p_i(r)=N_i(r\geq r_0)/N_i^{\mathrm{total}}$,

where $N_i(r\geq r_0)$ is the number of type-$i$ photons in the Crab cluster beyond $r_0$, and $N_i^{\mathrm{total}}$ is the total number of the type-$i$ photons in the Crab cluster.

\item \textbf{Likelihood Analysis:}
For a test cluster with $N_0$ front and $N_1$ back photons, we compute the log-likelihood value:
\begin{equation}
\mathcal{L}=-\left[\sum_{j=1}^{N_0} \ln{p_0(r_j)}+\sum_{k=1}^{N_1} \ln{p_1(r_k)}\right]
\label{eq:llh}
\end{equation}

To assess significance, we generate a likelihood distribution by randomly drawing 10,000 mock samples of $N_0$ and $N_1$ photons from the Crab cluster and calculating $\mathcal{L}$ for each.
\item \textbf{Extension Significance:}
The p-value for a point source is:
$P_{\mathrm{ps}}=N(\mathcal{L}>\mathcal{L}_0)/N_{\mathrm{tot}}$, where $\mathcal{L}_0$ is the likehood value for the analyzed cluster, $N(\mathcal{L}>\mathcal{L}_0)$ is the number of mock samples from the Crab distribution with likelihoods exceeding $\mathcal{L}_0$, and $N_{\mathrm{tot}}=10,000$ is the total number of mock samples. A low $P_{\mathrm{ps}}<0.01$ suggests significant extension.
\end{enumerate}

\subsection{Transients} 
While Section~\ref{subsec:dbscan} focused on spatial clustering, we now incorporate temporal constraints by requiring photon arrival times to lie within a 3-day window.  This modification accounts for the transient nature of potential sources. We chose this temporal scale in order  to select short  and  bright VHE flares, which have a typical duration of days.

For temporal clustering, the background rate $\lambda^*$ is adjusted to reflect the reduced time window:
\begin{equation}
\lambda^*=\lambda^{'}\frac{\delta t_{\mathrm{win}}}{T_{\mathrm{obs}}}
\label{eq:lambda_transient}
\end{equation}
where $\delta t_{\mathrm{win}}=3$~days  and $T_{\mathrm{obs}}\approx 16$~years is the total observation time. This scaling gives the expected background count within a 3-day interval.
Using Equation~(\ref{eq:pglob}) with $\lambda^*$, the revised threshold for cluster detection becomes $K=2$ (a 'doublet'), as even two temporally coincident photons are statistically significant at this low  level of background. The significance ($P_{\mathrm{glob}}$) of each candidate transient is recalculated using the methodology outlined in Section~\ref{subsec:dbscan}, now incorporating both spatial and temporal proximity.

\section{Results}
\label{sec:results}
\subsection{Clusters}

Our analysis identified 107 clusters with a global significance threshold of $P_{\mathrm{glob}}<0.05$. All of these clusters have an associated source from the Fourth Fermi Large Area Telescope catalog \citep[4FGL;][]{4FGL}. To prioritize novel VHE emitters, we imposed a stringent angular separation criterion, retaining only clusters located $>1^{\circ}$ from any source in the Third Catalog of Fermi-LAT High-Energy Sources \citep[3FHL;][]{3FHL} or TeVCat\footnote{\url{https://www.tevcat.org/}} \citep{TeVCat} (online interactive catalog for VHE sources).This produced three candidate VHE sources (Table~\ref{tab:vheclusters}) with no prior associations from VHE catalogs (3FHL and TeVCat respectively).

\begin{figure}[h]
                \centering
                \includegraphics[width=\linewidth]{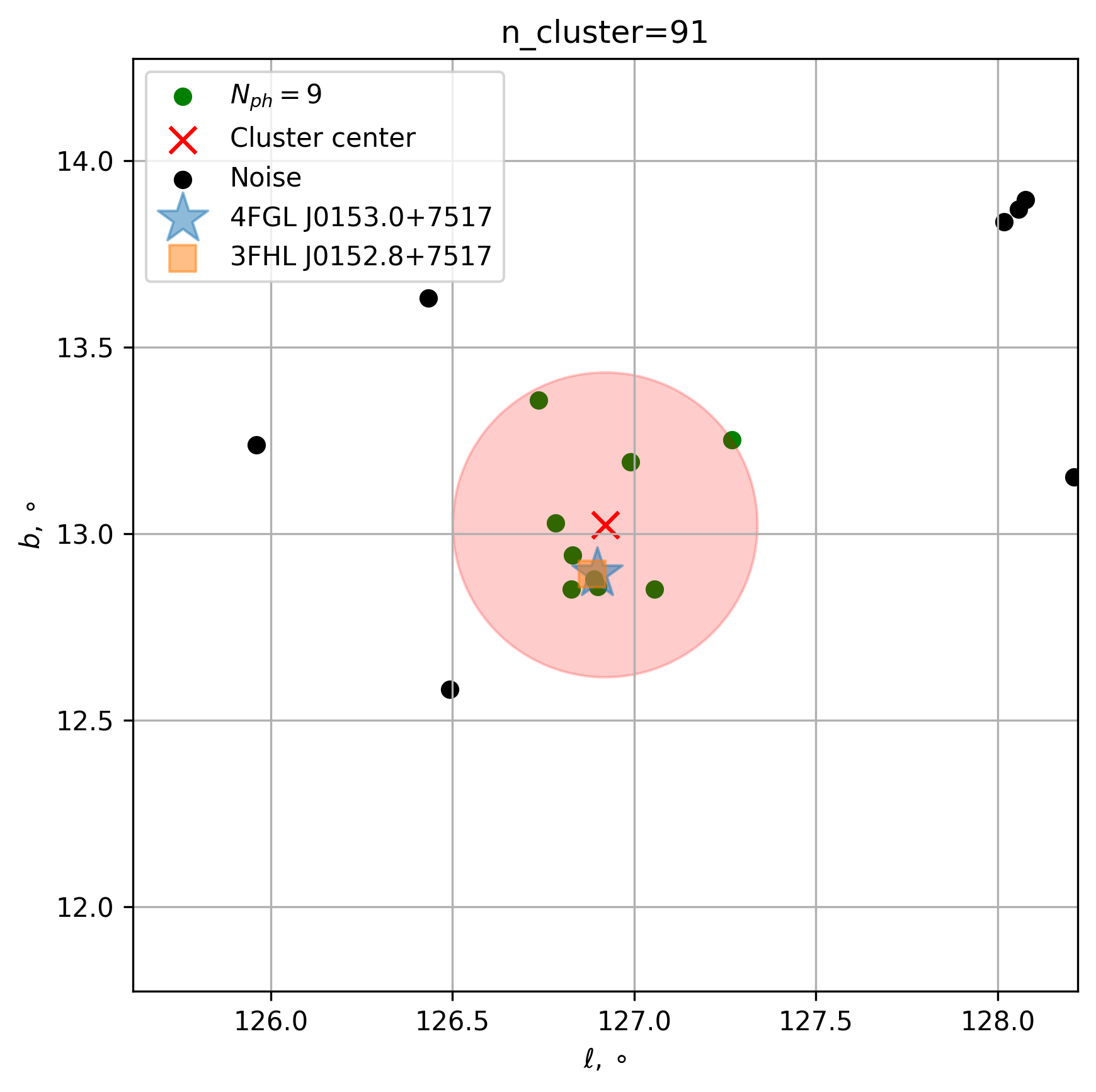}
                \caption{
                Spatial distribution of $\gamma$-ray photons  associated with the source 4FGL J0153.0+7517. The cluster's centroid is indicated by a red cross symbol, corresponding to Galactic coordinates $l=126.92^\circ, ~b=13.02^\circ$}
    \label{fig:j0153}
            \end{figure}

Most clusters showed no evidence of spatial extension, with photon distributions consistent with the Fermi-LAT PSF ($P_{\mathrm{ps}}>0.01$). The sole exception is Cluster~\# 91 ($N_{\mathrm{ph}}=9,~ P_{\mathrm{glob}}=2.54\times10^{-4}$, centered at $l=126.92^\circ, ~b=13.02^\circ$ and associated with 4FGL J0153.0+7517 (Fig.~\ref{fig:j0153})). This cluster exhibits marginal extension, with $P_{\mathrm{ps}}=6.8\times10^{-3}$ when referenced to its centroid and $P_{\mathrm{ps}}^\mathrm{3FHL}=0.033$  when anchored to the nearest 3FHL source position.

Notably, our method successfully recovered known extended sources, including the supernova remnant CTA 1 and the Large Magellanic Cloud (LMC), validating its sensitivity to resolved structures.

\subsubsection{4FGL J0039.1-2219}

\begin{figure}[h]
    
\begin{minipage}{\linewidth}
                \centering
                \includegraphics[width=\linewidth]{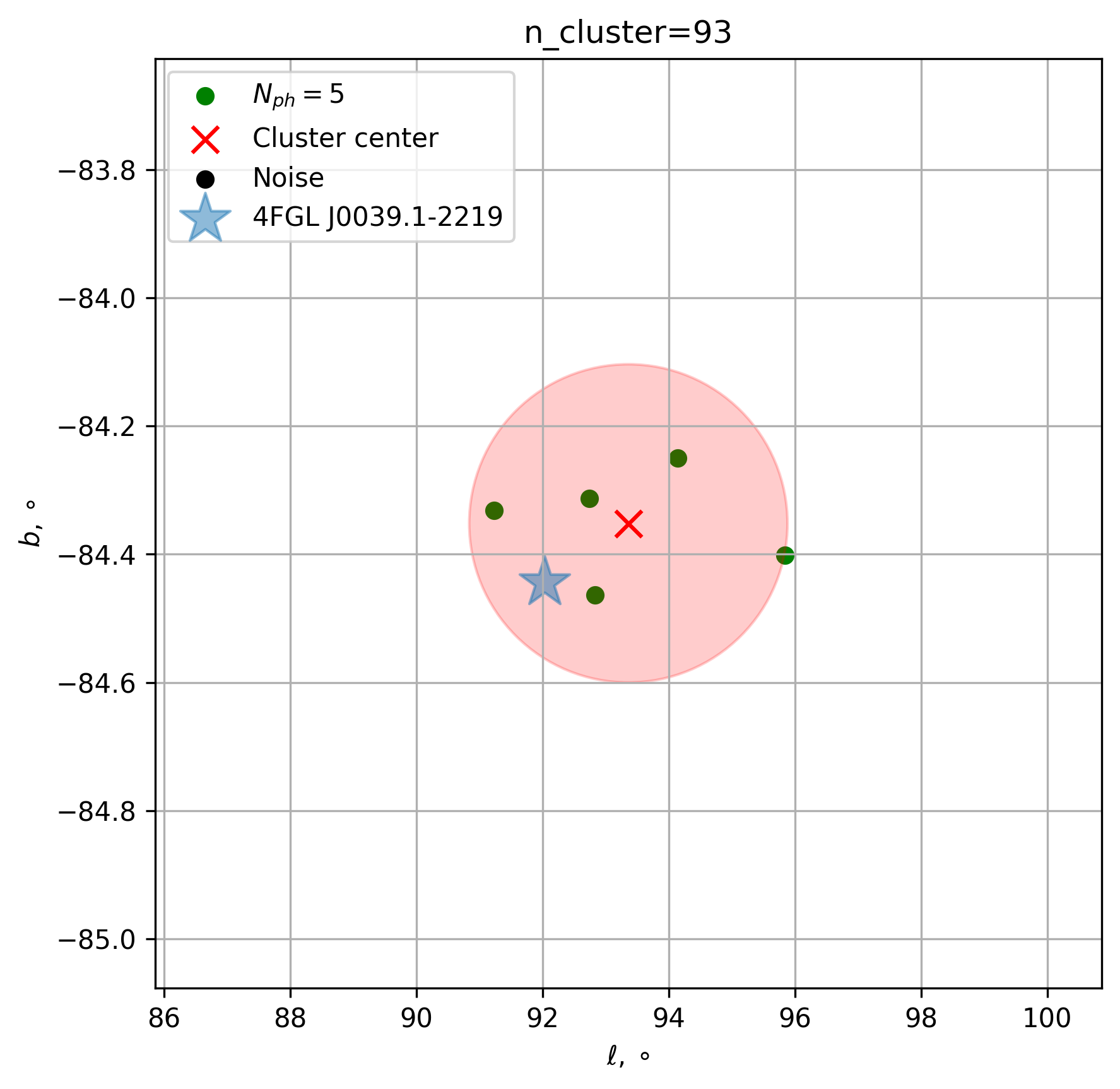}
                \end{minipage}
                \caption{
                Spatial distribution of $\gamma$-ray photons  associated with the source 4FGL J0039.1-2219. The cluster's centroid is indicated by a red cross symbol, corresponding to Galactic coordinates  	$l=93.36^\circ, ~b=-84.35^\circ$}
    \label{fig:j0039}
             
\end{figure}

This source is a low-redshift ($z=0.064$), low-luminosity BL Lacertae (BL Lac) blazar included in the 3HSP catalog \citep{3HSP} -- a comprehensive compilation of very high-energy (VHE) candidate sources selected using a multi-frequency dataset. The observed VHE flux is consistent  with the power-law index and flux extrapolated from lower-energy measurements reported in 4FGL. The angular distribution of $\gamma$-ray events around the source is shown in Fig.~\ref{fig:j0039}.

\subsubsection{4FGL J1544.3-0649}
\begin{figure}[h!]
                \centering
                \includegraphics[width=\linewidth]{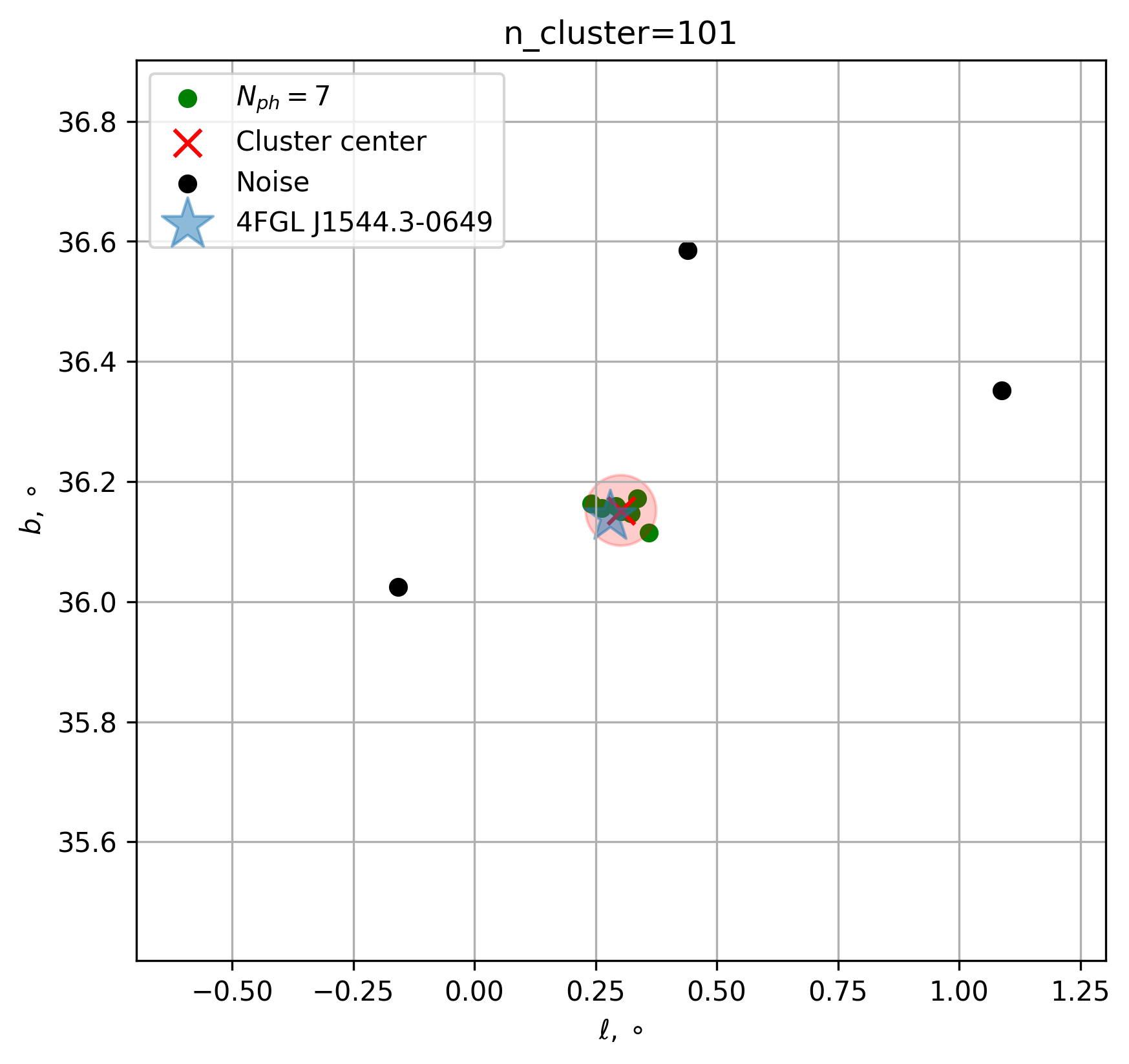}
                \caption{Spatial distribution of $\gamma$-ray photons  associated with the source 4FGL J1544.3-0649. The cluster's centroid is indicated by a red cross symbol, corresponding to Galactic coordinates  	$l=0.30^\circ, ~b=36.15^\circ$
                }
    \label{fig:j1544}
            \end{figure}

4FGL J1544.3-0649 (Fig.~\ref{fig:j1544}) is a blazar candidate first detected following a major $\gamma$-ray flare in May 2017 \citep{Bruni2018}. The source may represent a rare transient blazar with episodic high-energy (HE) and VHE emission and an exceptionally low duty cycle \citep{Sahakyan2021}. Alternatively, it could be an AGN exhibiting intermittent emission from secondary mini-jets whose orientations are misaligned relative to the primary jet axis \citep{Shao2022}.  It is evident that this source passed our cuts only because of its transient nature, which prevented its entering into the 3FHL, TeVCat and 3HSP catalogs as they were compiled prior to its flare.

\subsubsection{4FGL J0212.2-0219}

\begin{figure}[h!]
                \centering
                \includegraphics[width=\linewidth]{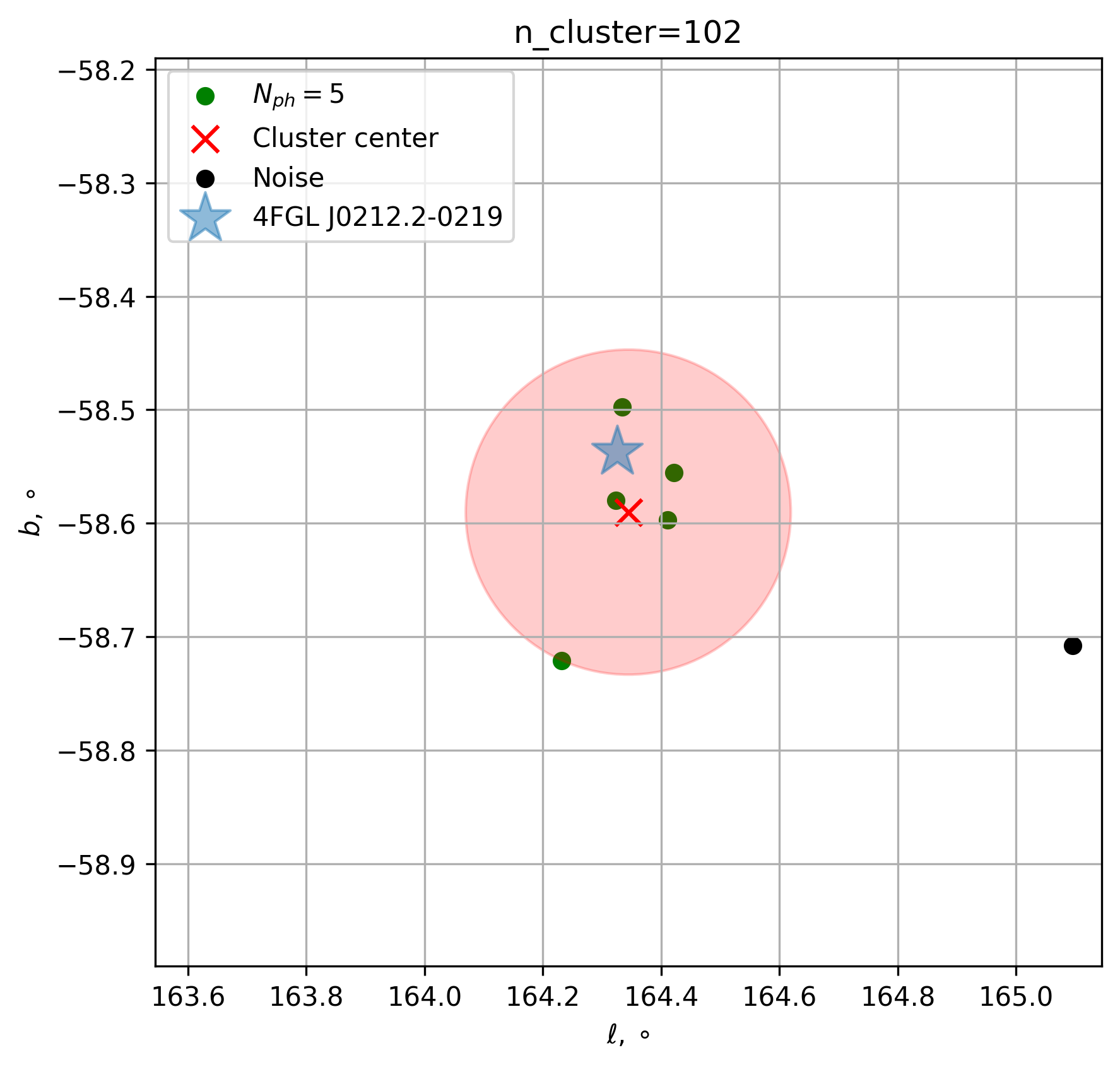}
                \caption{
                Spatial distribution of $\gamma$-ray photons  associated with the source 4FGL J0212.2-0219. The cluster's centroid is indicated by a red cross symbol, corresponding to Galactic coordinates  	$l=164.34^\circ, ~b=-58.59^\circ$
                }
    \label{fig:j0212}
            \end{figure}

This blazar (Fig.~\ref{fig:j0212}), located at $z=0.25$, is cataloged in the 3HSP \citep{3HSP} and was previously analyzed as a candidate for Imaging Atmospheric Cherenkov Telescopes (IACTs) in \citep{NievasRosillo2022}. While deemed too faint for detection by next-generation ground-based telescopes (e.g., CTA), our analysis reveals a marginal excess of very-high-energy (VHE) photons compared to predictions based on the spectral parameters reported in the 4FGL catalog. This discrepancy suggests the source's VHE flux may deviate from the extrapolated power-law spectrum, leaving open the possibility of detectability under improved sensitivity or during flaring episodes.

\subsection{Transients}
Blazars exhibit significant variability across their electromagnetic spectra, also in the VHE range. While Fermi-LAT data have been used to study VHE variability in bright sources \citep{Paliya2024}, we focus on extreme, short-term variability (flares) in weaker sources where traditional light-curve analysis is impractical due to low photon counts. Such flares could manifest as spatially and temporally clustered VHE photons, where even doublets (two photons within a few days) can achieve high statistical significance given the low background at VHE.
We identified 114 doublets of events  separated by not more than 3 days (259200 seconds) and 0.3 degrees. Most of them originate from bright sources, such as Mrk 421 (57 doublets), Mrk 501 (16), 1ES 1959+650 (11), BL Lac (5), 1ES 1218+304 (4), PKS 2155-304 (4), PG 1553+113  (3),  3FHL J0508.0+6737\footnote{A designation in the 3FHL catalog is given when there is no TeVCat association} (2), 1ES 1011+496 (1),  1ES 1727+502 (1),  3FHL J1457.8-4642 (1), 3FHL J0308.4+0408 (1), 3FHL J0847.0-2337 (1), 3FHL J1203.3-3924 (1), 3FHL J1037.6+5711 (1),  3C 66A (1),   PKS 1440-389 (1), PKS 0447-439 (1),  S5 0716+714 (1),  and one orphan flare without counterparts in the 4FGL, 3FHL and TeVCat catalogs. 
Additionally, we detected 12 triplets (10 from Mrk 421, 2 from 1ES 1959+650) and 1 quadruplet (Mrk 421).
We focused on the 6 doublets from sources without TeVCat associations or prior VHE flare reports (excluding 3FHL J0508.0+6737, which exhibited a VERITAS-detected flare in 2009 \citep{ATel_VERITAS_2009}). These doublets show exceptional significance ($P_{\mathrm{glob}} \in (10^{-8},~10^{-4})$), even without accounting for the spatial coincidence with 3FHL sources (Table~\ref{tab:transient_sources}).

Given the low statistics, two photons constrain the flare's expected event count to $0.35<N_{flare}<6.5$ (95\% CL). This translates to the VHE flux limits:
$0.17F_{VHE}^0<F_{VHE}<3.25F_{VHE}^0$, where $F_{VHE}^0=2/\mathcal{E}$ and $\mathcal{E}$ is the Fermi-LAT exposure during the doublet's time window.
For each candidate flare, we analyzed all VHE photons within  $0.3^{\circ}$ of  the associated source and  and  10-100 GeV photons\footnote{We will refer to them as HE photons in the following.} in this circle in the time interval ($t_0-10^7$~s, $t_0+10^7$~s), where $t_0$ is the doublet's midpoint.

\subsubsection{3FHL J0308.4+0408 (NGC 1218)}
The VHE  and HE   photon lists for this source are provided in Appendix Tables~\ref{tab:0308_vhe} and \ref{tab:0308_10GeV}. Only two VHE photons were detected within $0.3^{\circ}$ of the source over 16 years, implying a flare with a flux increase of $>10^{3}$ relative to the quiescent VHE level. Notably, only one of 36 HE photons coincided with the flare epoch (observed 5.5  Ms prior to the VHE event), suggesting an orphan VHE flare without a HE counterpart.

NGC 1218 is a S0 galaxy hosting the radio-bright Seyfert Type I nucleus 3C 78. While Seyfert galaxies sometimes exhibit GeV flaring activity \citep{Gokus2021}, VHE flares have been reported only in the Type II Seyfert NGC 1275 (Perseus Cluster core; \citep{NGC1275_MAGIC, NGC1275_LHAASO}). Intriguingly, recent IceCube observations associate Seyferts like NGC 1068 and NGC 7469 with astrophysical neutrinos \citep{NGC1068_neutrino, NGC7469_neutrino}, raising questions about their potential as transient multi-messenger sources.

\subsubsection{3FHL  J0847.0-2337 (PMN J0847-2337)}
PMN J0847-2337 is a candidate blazar exhibiting an unusual temporal pattern: a GeV-flare detected after a VHE one. Analysis of its photon distribution (Tables~\ref{tab:0847_vhe}, \ref{tab:0847_10GeV}) reveals that 8 of 90 HE  photons arrived within a 3 Ms  window starting 4 Ms  after the VHE flare, with 5 photons concentrated in a 10-day sub-interval.
Naturally, there is a strong observational bias against detection of such flares with IACTs, which usually  exploit HE flares  as a sort of external triggers.

\subsubsection{3FHL  J1037.6+5711 (GB6J1037+5711)}
This relatively bright BL Lac object exhibits 10 VHE and 325 HE photons in total. The VHE flare coincides with a prolonged period of enhanced HE activity (Tables~\ref{tab:1037_vhe}, \ref{tab:1037_10GeV}), and a transient event was independently detected by HAWC approximately one week after the Fermi-LAT doublet \citep{J1037_HAWC_ATel}.

Despite its brightness, this source lacks a secure redshift measurement  (Table~\ref{tab:transient_sources}). Spectroscopic observations in  SDSS DR7 reported $z=0.83\pm0.0004$  but with low confidence ($z_{\mathrm{conf}}=0.07$) \citep{SDSS7}. Follow-up observations with larger telescopes revealed a featureless optical spectrum, yielding only a lower limit of $z>0.25$  \citep{Paiano2020, Kasai2023}.
Photometric estimates from broadband SED modeling   gave a less reliable but intriguing value of $z=1.14$ \citep{Foschini2022}.

The source's AGN-dominated emission ($\sim0.5$~mag brightening over two decades) further obscures host galaxy features, complicating the redshift determination.

\subsubsection{3FHL J1203.3-3924 (PMN J1203-3926)}
This source is a blazar candidate, possibly of the BL Lac type. It exhibited a VHE flare without a HE counterpart, and none of the 40 HE photons arrived in the 20 Ms interval of interest (see Table \ref{tab:1203_vhe}).

\subsubsection{3FHL J1457.8-4642 (PMN J1457-4642)}

Only two VHE photons were detected from this blazar candidate during the entire 16-year  time span of the observations, both constituting a temporally clustered doublet. This VHE activity was preceded by enhanced HE emission: 4 out of the 12 HE photons arrived within a 1.5 Ms  window starting 3 Ms  prior to the VHE event (Tables \ref{tab:1457_vhe}, \ref{tab:1457_10GeV}).

This HE precursor to VHE flaring is a common feature of blazar variability. For instance, during the 2011 outburst of the BL Lacertae, HE activity preceded VHE emission by $\sim$ month \citep{Arlen2013}.

\subsubsection{NGC 5549?}
The lack of a 3FHL/4FGL counterpart complicates the identification of the host source for this VHE doublet. While several galaxies reside within the same galaxy group as NGC 5549 -- a low-luminosity AGN exhibiting faint nuclear activity -- NGC 5549 itself emerges as the most plausible candidate due to its weak but discernible AGN signatures.

The $0.2^{\circ}$ region around the doublet's centroid contains only distant quasars ($z>0.75$), which are unlikely counterparts given their extreme redshifts and spectral characteristics. Strikingly, not only  VHE flare (see Table \ref{tab:ngc5549_vhe}) lacks any HE counterpart: zero HE photons were detected in $0.3^{\circ}$ in all 16 years of observations.

\subsubsection{Energetics and spectral properties  of the flares}
The VHE  luminosities of the flaring sources exhibit extraordinary variability, increasing by up to three orders of magnitude compared to their quiescent states. During peak activity, some flares reached VHE(!) luminosities of $\mathcal{O}(10^{47}~\mathrm{erg ~s^{-1}})$, rivaling the bolometric output of luminous quasars.
For context, we estimate the VHE luminosities of two well-studied blazars during their brightest 3-day intervals -- Mrk 421, $1.7\times10^{46}~\mathrm{erg~s^{-1}}$ (MET 240723290--240962846; 18--20 August 2008, 4 photons), and     BL Lacertae:  $6.0\times10^{45}~\mathrm{erg~s^{-1}}$  (MET 647619845 -- 647750055; 10-- 12 July 2021, 2 photons).
These values are an order of magnitude lower than the most extreme flares identified in this work, underscoring the exceptional nature of the newly detected events.
Several flares lacked detectable HE  counterparts, implying extremely hard spectra. This is corroborated by the energies of VHE photons in key doublets:  3FHL J1203.3-3924 (172 GeV and 203 GeV photons), NGC 5549 (orphan flare;  156 GeV and 593 GeV photons).

An interesting representative of this class could be  the VHE quadruplet from Mrk 421 (4 photons within 3 days), which occurred without contemporaneous HE or X-ray activity \citep{Abdo2011_Mrk421, Bartoli2016}. This event presents two hypotheses, either there was   a missed HE precursor --  a hypothetical HE flare $\sim 1$ month prior could have escaped detection due to Fermi-LAT's commissioning phase at the time, or it was just a statistical fluctuation. Given that in 5840 days there were observed 372 VHE photons from this source, the probability of a random 4-photon cluster  with a 180 ks duration is $9\%$ (Poissonian estimate).

The total  absence of HE emission could pose a challenge to the standard synchrotron self-Compton (SSC) model, suggesting that some alternative mechanisms of emission could be responsible for such flares.

Detected flares bear striking resemblance to several instances of  fast variability of another misaligned sources, i.e. IC 310 and M 87 \citep{Rieger2018,Rieger2019}: in both cases we need some new  additional VHE component to account for the observed spectrum. Still, if we took quiescent HE luminosity as a proxy for a jet power $L_j$ \citep{Nemmen2012}, VHE -jet luminosity ratios $L_{\mathrm{VHE}}/L_{j}$ have considerably higher values for our set, which possibly disfavor models with magnetospheric origin of the emission and various jet-object interaction scenarios.


\section{Summary and conclusions}
\label{sec:conclusions}

In this paper we searched for spatio-temporal clusters of the VHE events ($E>100$~GeV) in 16 years of the Fermi-LAT data at galactic latitudes $|b|>10^{\circ}$. Using the DBSCAN algorithm,  we found 107 clusters. All, but three of them,  were  associated with already known VHE sources, listed in the 3FHL or TeVCat catalogs (Table \ref{tab:clusters_full_list}). From remaining three (Table \ref{tab:vheclusters}), 4FGL J1544.3-0649, which had an epoch of activity in 2017-2018, could belong to a rare class of transient blazars and because of that evaded inclusion into VHE catalogs. The other sources, 4FGL J0039.1-2219 and 4FGL J0039.1-2219 are weak BL Lacs, that could be a tip of an iceberg of much wider population of weaker sources, that would be detected with the next-generation IACTs. 

We also searched for flaring  VHE activity, looking for event multiplets within  short temporal intervals of 3 days. We found 114 doublets, most of them (57) from Mrk 421 -- the brightest  extragalactic source with a very high level of variability.
We focused on the 6 doublets without TeVCat sources in their proximity. Five of them were associated with 3FHL sources (J0308.4+0408, J0847.0-2337, J1037.6+5711, J1203.3-3924, J1457.8-4642), and one has no association with any known HE/VHE sources (see Table \ref{tab:transient_sources}).

Due to the limited effective area of the Fermi-LAT this approach is sensitive  only to the brightest flares with a very high luminosity that in cases of 3FHL J1037.6+5711 and 3FHL J1457.8-4642 could reach $\mathcal{O}(10^{47}~\mathrm{erg~ s^{-1}})$ (isotropic), which is, e.g., an order of magnitude larger than the brightest BL Lac flare in July 2021.

One of the sources -- 3FHL J0308.4+0408(NGC 1218) -- is a Seyfert Type I galaxy. No objects of this type not belonging to narrow-line Sy 1 subclass were known as VHE sources before. A fortiori, no short energetic VHE flares were anticipated from such objects. 
Even more striking is the plausible association of the 'orphan' doublet with the LINER galaxy NGC 5549. This type of galaxies are considered to harbor low-luminosity AGNs in their centers  and currently they are below sensitivity limit of HE/VHE telescopes.  Still, there was a major extremely hard flare, during which the VHE flux ($E>100$~GeV) exceeded the average flux at HE ($E>10$~GeV) at least a thousandfold, and the VHE luminosity of the AGN surpassed $10^{45}~\mathrm{erg~s^{-1}}$.  It is worth noting, that non-blazar AGNs are a small minority in the full population of TeV sources,  with only 4 Fanaroff-Riley I galaxies present in the TeVCat (M87, 3C 264, NGC 1275, and Cen A), so a detection of VHE flares from low-luminous non-aligned AGNs could have considerable significance for the field of  AGN jet  studies.

HE behaviour during these flares was quite diverse: 3FHL J1037.6+5711 had coinciding VHE/HE flares, the HE flare in 3FHL J1457.8-4642  preceded the VHE one by about a month, there was HE activity 1.5 month \textit{after} the VHE flare in 3FHL J0847.0-2337, which is quite uncommon, and in 3FHL J0308.4+0408, 3FHL J1203.3-3924, NGC 5549? there were no HE counterparts whatsoever.


\textit{Note:} When this paper was in a final stage of preparation, another work in  which a DBSCAN algorithm was used to search for clusters in GeV Fermi data, appeared \citep{Malyshev2025}. Their analysis focused on identifying extended sources in the Galactic plane, contrasting with our emphasis on extragalactic transients and high-latitude sources.

\section*{Acknowledgements}
The study was funded by a grant from the Russian Science Foundation 25-12-00111 (https://rscf.ru/en/project/25-12-00111/). This research has made use of NASA's Astrophysics Data System and  the SIMBAD database,
operated at CDS, Strasbourg, France.
\bibliographystyle{elsarticle-harv-et-al}    
\bibliography{example}

\appendix

\onecolumn 
\section{Table of clusters}
\vspace{10pt}

\setlength{\LTpre}{-10pt}
\begin{longtable}{c r r c c l c l c}

\toprule
\multicolumn{1}{c}{\#} & 
\multicolumn{1}{c}{$l$, $^\circ$} & 
\multicolumn{1}{c}{$b$, $^\circ$} & 
\multicolumn{1}{c}{$N_{\text{ph}}$} & 
\multicolumn{1}{c}{$P_{\text{glob}}$} & 
\multicolumn{1}{c}{Nearest 4FGL} & 
\multicolumn{1}{c}{$\rho_{\text{4FGL}},\ ^\circ$} & 
\multicolumn{1}{c}{Nearest 3FHL} & 
\multicolumn{1}{c}{$\rho_{\text{3FHL}},\ ^\circ$} \\

\midrule
\endfirsthead

\toprule
\multicolumn{1}{c}{\#} & 
\multicolumn{1}{c}{$l$, $^\circ$} & 
\multicolumn{1}{c}{$b$, $^\circ$} & 
\multicolumn{1}{c}{$N_{\text{ph}}$} & 
\multicolumn{1}{c}{$P_{\text{glob}}$} & 
\multicolumn{1}{c}{Nearest 4FGL} & 
\multicolumn{1}{c}{$\rho_{\text{4FGL}},\ ^\circ$} & 
\multicolumn{1}{c}{Nearest 3FHL} & 
\multicolumn{1}{c}{$\rho_{\text{3FHL}},\ ^\circ$} \\

\midrule
\endhead
\midrule
\multicolumn{8}{r}{{Continued on next page}} \\
\midrule
\endfoot
\bottomrule
\endlastfoot
0 & 179.84 & 65.04 & 372 & --- & J1104.4+3812 & 0.00715 & J1104.4+3812 & 0.00496 \\
1 & 76.96 & 21.77 & 10 & $3.46 \times 10^{-11}$ & J1838.8+4802 & 0.05751 & J1838.8+4802 & 0.05598 \\
2 & 165.06 & -31.75 & 7 & $6.35 \times 10^{-06}$ & J0319.8+1845 & 0.07170 & J0319.8+1845 & 0.07020 \\
3 & 17.71 & -52.27 & 87 & $2.45 \times 10^{-166}$ & J2158.8-3013 & 0.02860 & J2158.8-3013 & 0.02726 \\
4 & 102.90 & -24.76 & 6 & $3.25 \times 10^{-05}$ & J2322.7+3436 & 0.03764 & J2322.6+3436 & 0.03232 \\
5 & 98.32 & -18.50 & 7 & $2.78 \times 10^{-05}$ & J2250.0+3825 & 0.08845 & J2250.0+3825 & 0.08653 \\
6 & 298.75 & -15.82 & 6 & $3.49 \times 10^{-03}$ & J1130.5-7801 & 0.09545 & J1130.5-7801 & 0.09011 \\
7 & 150.23 & -13.73 & 6 & $1.28 \times 10^{-05}$ & J0316.8+4120 & 0.05194 & J0316.6+4120 & 0.05965 \\
8 & 18.22 & -14.33 & 14 & $1.11 \times 10^{-17}$ & J1917.7-1921 & 0.02275 & J1917.7-1921 & 0.01938 \\
9 & 29.48 & 68.20 & 35 & $1.16 \times 10^{-65}$ & J1427.0+2348 & 0.00959 & J1427.0+2348 & 0.01108 \\
10 & 63.59 & 38.85 & 124 & $2.24 \times 10^{-259}$ & J1653.8+3945 & 0.00947 & J1653.8+3945 & 0.00883 \\
11 & 143.79 & 15.94 & 42 & $3.83 \times 10^{-44}$ & J0507.9+6737 & 0.04702 & J0508.0+6737 & 0.04749 \\
12 & 195.36 & -19.61 & 7 & $2.19 \times 10^{-08}$ & J0509.4+0542 & 0.04642 & J0509.4+0542 & 0.04134 \\
13 & 191.78 & -33.18 & 8 & $3.89 \times 10^{-08}$ & J0416.9+0105 & 0.04268 & J0416.8+0105 & 0.02835 \\
14 & 282.74 & -40.76 & 5 & $9.84 \times 10^{-05}$ & J0353.0-6831 & 0.04569 & J0353.0-6832 & 0.05355 \\
15 & 248.79 & -39.93 & 43 & $3.59 \times 10^{-85}$ & J0449.4-4350 & 0.01883 & J0449.4-4350 & 0.02069 \\
16 & 21.91 & 44.00 & 99 & $1.64 \times 10^{-167}$ & J1555.7+1111 & 0.03366 & J1555.7+1111 & 0.03280 \\
17\footnote{This clusters contains 2 close sources: J1221.3+3010 (44 photons) and J1217.9+3006 (17 photons)} & 187.06 & 82.53 & 61 & $5.93 \times 10^{-124}$ & J1221.3+3010 & 0.23119 & J1221.3+3010 & 0.22455 \\
18 & 77.35 & 64.88 & 11 & $1.40 \times 10^{-17}$ & J1428.5+4240 & 0.06910 & J1428.5+4240 & 0.06110 \\
19 & 98.03 & 17.67 & 152 & $1.15 \times 10^{-236}$ & J2000.0+6508 & 0.02509 & J1959.9+6508 & 0.02636 \\
20 & 306.04 & 19.78 & 5 & $4.71 \times 10^{-02}$ & J1307.6-4259 & 0.03644 & J1307.6-4259 & 0.03435 \\
21 & 165.54 & 52.72 & 36 & $1.29 \times 10^{-63}$ & J1015.0+4926 & 0.01263 & J1015.0+4926 & 0.01314 \\
22 & 3.26 & 10.59 & 7 & $9.76 \times 10^{-04}$ & J1714.0-2029 & 0.06513 & J1714.0-2028 & 0.06933 \\
23 & 249.67 & 18.49 & 12 & $6.73 \times 10^{-12}$ & J0912.9-2102 & 0.09428 & J0912.9-2103 & 0.08537 \\
24 & 350.39 & -32.59 & 23 & $9.21 \times 10^{-27}$ & J2009.4-4849 & 0.01434 & J2009.4-4849 & 0.00996 \\
25 & 190.27 & 10.96 & 36 & $2.00 \times 10^{-57}$ & J0650.7+2503 & 0.03895 & J0650.7+2503 & 0.03922 \\
26 & 56.78 & 11.92 & 5 & $8.36 \times 10^{-04}$ & J1850.5+2631 & 0.02421 & J1850.4+2631 & 0.04292 \\
27 & 166.27 & 32.91 & 19 & $4.81 \times 10^{-37}$ & J0809.8+5218 & 0.02201 & J0809.8+5218 & 0.01780 \\
28 & 11.02 & -26.72 & 6 & $2.53 \times 10^{-03}$ & J1958.3-3010 & 0.08546 & J1958.3-3011 & 0.09158 \\
29 & 247.77 & 12.19 & 11 & $8.34 \times 10^{-14}$ & J0847.0-2336 & 0.01326 & J0847.0-2337 & 0.01002 \\
30 & 131.91 & 45.63 & 13 & $1.97 \times 10^{-26}$ & J1136.4+7009 & 0.01486 & J1136.5+7009 & 0.01732 \\
31 & 320.77 & -11.15 & 5 & $3.40 \times 10^{-05}$ & J1610.7-6648 & 0.02619 & J1610.6-6649 & 0.01945 \\
32 & 278.33 & -60.78 & 12 & $1.25 \times 10^{-17}$ & J0209.3-5228 & 0.01143 & J0209.3-5229 & 0.01831 \\
33 & 225.46 & 67.36 & 8 & $1.39 \times 10^{-11}$ & J1117.0+2013 & 0.05154 & J1117.0+2014 & 0.04253 \\
34 & 144.03 & 28.04 & 24 & $7.76 \times 10^{-36}$ & J0721.9+7120 & 0.04577 & J0721.8+7120 & 0.04924 \\
35 & 93.96 & -81.35 & 5 & $2.93 \times 10^{-06}$ & J0033.5-1921 & 0.13761 & J0033.5-1921 & 0.13991 \\
36 & 140.17 & -16.79 & 30 & $4.37 \times 10^{-36}$ & J0222.6+4302 & 0.02852 & J0222.6+4302 & 0.02812 \\
37 & 112.89 & -10.07 & 6 & $2.09 \times 10^{-02}$ & J2347.0+5141 & 0.15451 & J2347.0+5142 & 0.15805 \\
38 & 214.61 & -60.20 & 10 & $1.48 \times 10^{-20}$ & J0303.4-2407 & 0.02414 & J0303.4-2407 & 0.02177 \\
39 & 309.50 & 19.42 & 9 & $6.77 \times 10^{-07}$ & J1325.5-4300 & 0.03100 & J1325.5-4300 & 0.02866 \\
40 & 243.45 & -19.99 & 8 & $3.16 \times 10^{-07}$ & J0627.0-3529 & 0.02001 & J0627.1-3528 & 0.02425 \\
41 & 60.26 & 63.68 & 5 & $2.65 \times 10^{-09}$ & J1448.0+3608 & 0.03231 & J1447.9+3608 & 0.04511 \\
42 & 261.35 & -11.34 & 5 & $4.08 \times 10^{-07}$ & J0746.6-4754 & 0.03600 & J0746.6-4755 & 0.02292 \\
43 & 132.46 & -22.96 & 8 & $1.38 \times 10^{-07}$ & J0136.5+3906 & 0.03699 & J0136.5+3906 & 0.03962 \\
44 & 266.93 & 20.05 & 6 & $1.73 \times 10^{-10}$ & J1010.2-3119 & 0.01853 & J1010.2-3119 & 0.02442 \\
45 & 157.46 & 25.40 & 5 & $2.58 \times 10^{-04}$ & J0710.4+5908 & 0.05572 & J0710.4+5908 & 0.06219 \\
46 & 350.09 & -67.56 & 6 & $5.71 \times 10^{-12}$ & J2324.7-4041 & 0.04388 & J2324.7-4040 & 0.04433 \\
47 & 351.14 & -26.99 & 5 & $2.88 \times 10^{-04}$ & J1936.9-4720 & 0.03668 & J1936.9-4720 & 0.02770 \\
48 & 240.67 & -43.58 & 5 & $2.33 \times 10^{-04}$ & J0428.6-3756 & 0.04557 & J0428.6-3756 & 0.04820 \\
49 & 152.42 & -18.41 & 6 & $7.11 \times 10^{-08}$ & J0312.9+3614 & 0.00644 & J0312.8+3614 & 0.01734 \\
50 & 138.92 & 30.81 & 18 & $3.39 \times 10^{-30}$ & J0805.4+7534 & 0.02880 & J0805.5+7534 & 0.02682 \\
51 & 263.52 & -31.48 & 24 & $1.25 \times 10^{-36}$ & J0543.9-5531 & 0.00615 & J0543.9-5532 & 0.00509 \\
52 & 123.75 & 58.79 & 6 & $2.56 \times 10^{-09}$ & J1248.3+5820 & 0.02066 & J1248.3+5820 & 0.01891 \\
53 & 278.42 & -53.12 & 8 & $2.44 \times 10^{-09}$ & J0244.6-5819 & 0.03614 & J0244.4-5819 & 0.03664 \\
54 & 150.58 & -13.28 & 15 & $2.72 \times 10^{-21}$ & J0319.8+4130 & 0.01751 & J0319.8+4130 & 0.01879 \\
55 & 47.45 & -30.22 & 5 & $4.84 \times 10^{-02}$ & J2104.3-0212 & 0.11744 & J2104.2-0212 & 0.11968 \\
56 & 107.05 & -61.87 & 5 & $5.24 \times 10^{-05}$ & J0022.0+0006 & 0.07363 & J0022.0+0006 & 0.05967 \\
57 & 307.51 & 20.06 & 7 & $4.08 \times 10^{-06}$ & J1315.0-4236 & 0.04673 & J1315.0-4237 & 0.05338 \\
58 & 161.42 & 54.42 & 9 & $3.11 \times 10^{-15}$ & J1031.3+5053 & 0.03815 & J1031.3+5053 & 0.03205 \\
59 & 67.76 & -10.79 & 6 & $4.35 \times 10^{-05}$ & J2042.1+2427 & 0.02653 & J2042.0+2428 & 0.03159 \\
60 & 167.84 & 66.11 & 15 & $1.07 \times 10^{-22}$ & J1120.8+4212 & 0.05440 & J1120.8+4212 & 0.05593 \\
61 & 279.56 & -31.72 & 22 & $2.00 \times 10^{-22}$ & J0537.8-6909 & 0.02272 & J0537.9-6909 & 0.01731 \\
62 & 119.73 & 10.81 & 8 & $8.05 \times 10^{-03}$ & J0007.0+7303 & 0.35210 & J0007.0+7303 & 0.34733 \\
63 & 13.09 & -78.09 & 5 & $2.09 \times 10^{-07}$ & J2359.0-3038 & 0.09019 & J2359.1-3038 & 0.09073 \\
64 & 283.89 & 74.51 & 7 & $3.37 \times 10^{-13}$ & J1230.8+1223 & 0.02985 & J1230.8+1223 & 0.03775 \\
65 & 347.88 & 24.23 & 8 & $4.08 \times 10^{-05}$ & J1548.8-2250 & 0.06035 & J1548.7-2250 & 0.06277 \\
66 & 340.69 & 27.48 & 9 & $1.89 \times 10^{-06}$ & J1517.7-2422 & 0.09283 & J1517.6-2422 & 0.10043 \\
67 & 201.92 & -45.70 & 5 & $1.31 \times 10^{-06}$ & J0349.4-1159 & 0.01386 & J0349.3-1159 & 0.01245 \\
68 & 153.96 & 14.23 & 5 & $4.49 \times 10^{-03}$ & J0540.5+5823 & 0.03701 & J0540.5+5823 & 0.04052 \\
69 & 325.66 & 18.72 & 20 & $1.30 \times 10^{-39}$ & J1443.9-3908 & 0.02015 & J1443.9-3908 & 0.01497 \\
70 & 145.70 & 43.10 & 5 & $5.60 \times 10^{-05}$ & J0958.7+6534 & 0.04408 & J0958.7+6533 & 0.04688 \\
71 & 133.51 & 80.54 & 5 & $3.91 \times 10^{-07}$ & J1243.2+3627 & 0.09365 & J1243.2+3627 & 0.08915 \\
72 & 92.60 & -10.44 & 44 & $1.41 \times 10^{-66}$ & J2202.7+4216 & 0.00625 & J2202.7+4216 & 0.01138 \\
73 & 233.80 & -17.58 & 7 & $1.07 \times 10^{-11}$ & J0622.3-2605 & 0.02166 & J0622.4-2606 & 0.03592 \\
74 & 100.77 & -13.23 & 7 & $1.49 \times 10^{-08}$ & J2247.8+4413 & 0.04598 & J2247.9+4413 & 0.04050 \\
75 & 201.89 & 18.13 & 5 & $8.21 \times 10^{-06}$ & J0738.1+1742 & 0.06056 & J0738.1+1742 & 0.06771 \\
76 & 351.25 & 40.15 & 7 & $7.07 \times 10^{-05}$ & J1512.8-0906 & 0.03401 & J1512.8-0906 & 0.02758 \\
77 & 34.19 & 24.50 & 11 & $3.01 \times 10^{-10}$ & J1725.0+1152 & 0.06263 & J1725.0+1152 & 0.06015 \\
78 & 77.11 & 33.55 & 14 & $1.70 \times 10^{-20}$ & J1728.3+5013 & 0.03547 & J1728.3+5013 & 0.04306 \\
79 & 119.75 & 17.80 & 7 & $5.76 \times 10^{-06}$ & J2340.8+8015 & 0.08647 & J2340.8+8015 & 0.08861 \\
80 & 18.62 & -21.00 & 8 & $1.43 \times 10^{-05}$ & J1944.9-2143 & 0.10907 & J1944.9-2143 & 0.10566 \\
81 & 67.32 & 13.40 & 7 & $2.33 \times 10^{-04}$ & J1904.1+3627 & 0.03917 & J1904.1+3627 & 0.04055 \\
82 & 229.58 & -66.30 & 6 & $7.19 \times 10^{-08}$ & J0238.4-3116 & 0.06258 & J0238.4-3117 & 0.06161 \\
83 & 87.38 & 73.46 & 6 & $2.32 \times 10^{-07}$ & J1341.2+3958 & 0.06589 & J1341.2+3959 & 0.05339 \\
84 & 166.05 & 27.27 & 5 & $3.50 \times 10^{-03}$ & J0733.4+5152 & 0.04548 & J0733.4+5152 & 0.04788 \\
85 & 327.77 & 14.64 & 8 & $7.36 \times 10^{-06}$ & J1503.6-4146 & 0.02214 & J1451.8-4145 & 2.18606 \\
86 & 282.17 & -61.32 & 5 & $2.12 \times 10^{-04}$ & J0156.9-5301 & 0.06521 & J0156.7-5302 & 0.07746 \\
87 & 293.36 & -17.56 & 5 & $3.28 \times 10^{-06}$ & J0953.4-7659 & 0.05297 & J0953.3-7659 & 0.04906 \\
88 & 33.37 & 70.59 & 6 & $6.57 \times 10^{-08}$ & J1417.9+2543 & 0.10932 & J1418.0+2543 & 0.11528 \\
89 & 213.44 & -83.44 & 6 & $1.49 \times 10^{-12}$ & J0120.4-2701 & 0.09762 & J0120.4-2701 & 0.10540 \\
90 & 325.24 & 19.34 & 7 & $6.09 \times 10^{-09}$ & J1440.6-3846 & 0.04781 & J1440.6-3846 & 0.04649 \\
91 & 126.92 & 13.02 & 9 & $2.54 \times 10^{-04}$ & J0153.0+7517 & 0.13220 & J0152.8+7517 & 0.13704 \\
92 & 134.12 & 39.25 & 6 & $1.08 \times 10^{-04}$ & J1031.1+7442 & 0.08638 & J1031.2+7442 & 0.07792 \\
93 & 93.36 & -84.35 & 5 & $7.91 \times 10^{-07}$ & J0039.1-2219 & 0.15912 & J0033.5-1921 & 3.13907 \\
94 & 95.90 & -52.36 & 6 & $1.23 \times 10^{-07}$ & J2346.7+0705 & 0.06576 & J2346.6+0705 & 0.05724 \\
95 & 276.86 & -32.66 & 5 & $4.08 \times 10^{-02}$ & J0531.8-6639e & 0.16534 & J0531.8-6639e & 0.16534 \\
96 & 100.56 & 30.83 & 5 & $1.57 \times 10^{-03}$ & J1748.6+7005 & 0.12346 & J1748.6+7006 & 0.12108 \\
97 & 358.32 & -48.23 & 6 & $3.34 \times 10^{-02}$ & J2139.4-4235 & 0.09523 & J2139.4-4234 & 0.10106 \\
98 & 151.77 & 51.78 & 10 & $1.19 \times 10^{-18}$ & J1037.7+5711 & 0.00843 & J1037.6+5711 & 0.01053 \\
99 & 107.43 & -26.10 & 5 & $2.70 \times 10^{-02}$ & J2343.6+3438 & 0.08011 & J2343.6+3439 & 0.06839 \\
100 & 129.94 & -37.22 & 6 & $7.57 \times 10^{-05}$ & J0115.8+2519 & 0.06736 & J0115.8+2519 & 0.06616 \\
101 & 0.30 & 36.15 & 7 & $2.48 \times 10^{-12}$ & J1544.3-0649 & 0.02103 & J1549.9-0659 & 1.41385 \\
102 & 164.34 & -58.59 & 5 & $2.98 \times 10^{-06}$ & J0212.2-0219 & 0.05389 & J0217.8+0143 & 4.33239 \\
103 & 323.26 & -34.37 & 5 & $9.16 \times 10^{-03}$ & J2040.2-7115 & 0.16504 & J2040.3-7116 & 0.17108 \\
104 & 93.31 & 19.74 & 6 & $6.45 \times 10^{-07}$ & J1926.8+6154 & 0.00103 & J1926.9+6154 & 0.00839 \\
105 & 152.90 & -36.74 & 6 & $1.57 \times 10^{-05}$ & J0232.8+2018 & 0.16848 & J0232.8+2017 & 0.16124 \\
106 & 8.58 & 59.99 & 5 & $3.29 \times 10^{-04}$ & J1442.7+1200 & 0.19075 & J1442.8+1200 & 0.20314 \\
\label{tab:clusters_full_list}
\end{longtable}

\section{Table of VHE cluster candidates}
\vspace{-5pt}
\begin{table}[h]
\setlength{\tabcolsep}{3pt}
    \centering
    \begin{tabular}{ccccccccc}
    \toprule
         \# & $l,^{\circ}$  & $b,^{\circ}$ & $N_{\mathrm{ph}}$ &$P_{\mathrm{glob}}$ & 4FGL name&$F_{1-100}$& $\gamma$ & $z$\\
         \hline
         93	&93.36 & -84.35	& 5 & $7.9 \times 10^{-7}$& J0039.1-2219&2.1&1.8&0.064\\
         101	&0.30  &  36.15 & 7 & $2.5 \times 10^{-12}$& J1544.3-0649&12.1&1.9&0.17\\
         102	&164.34& -58.59 & 5 & $3 \times 10^{-6}$ &J0212.2-0219&3.1&2.0&0.25\\
    \bottomrule
    \end{tabular}
\caption{Properties of unassociated VHE cluster candidates and their associations in 4FGL. Columns: (1) Cluster ID from Table~\ref{tab:clusters_full_list}; (2,3) Galactic longitude and latitude of cluster centroid; (4) Photon count; (5) Global probability of chance occurrence; (6) Associated 4FGL source; (7) Integrated GeV flux from 4FGL in $10^{-10}~\mathrm{cm^2~s^{-1}}$; (8) Spectral index; (9) Redshift.}
    \label{tab:vheclusters}
\end{table}

\section{Table of transients}
\vspace{-10pt}
\begin{table}[H]

    \begin{threeparttable}
    \caption{Properties of sources associated with transient doublet events. Columns show: (1) 3FHL designation; (2) angular separation from doublet center (degrees); (3) common source name; (4) redshift from \citep{Foschini2022} compilation (boldface indicates spectroscopic measurements, regular font photometric estimates); (5) source type (BLL = BL Lacertae, Sy 1 = Seyfert Type I, LIN = LINER galaxy); (6) 1--100 GeV flux from 4FGL in $10^{-9}~\mathrm{cm^{-2}~s^{-1}}$; (7) global p-value for transient detection; (8) doublet duration $\Delta t=t_2-t_1$ in $10^5~\mathrm{s}$; (9) Fermi-LAT exposure time in $10^{8}~\mathrm{cm^2~s}$; (10) observed VHE flux $F_{VHE}^0$ with 95\% CL (0.17--3.3)$\times F_{VHE}^0$ in $10^{-9}~\mathrm{cm^{-2}~s^{-1}}$; (11) derived luminosity in $10^{45}~\mathrm{erg~s^{-1}}$ .}
    
    \label{tab:transient_sources}
    \begin{tabular}{lllllccccccc}
    \toprule
        \multicolumn{1}{l}{3FHL name} & 
\multicolumn{1}{c}{$\rho$, $^\circ$} & 
\multicolumn{1}{l}{Assoc. name} & 
\multicolumn{1}{c}{$z$} & 
\multicolumn{1}{l}{type} & 
\multicolumn{1}{c}{$F_{1-100}$} & 
\multicolumn{1}{c}{$P_{\text{glob}}$} & 
\multicolumn{1}{c}{$\Delta t$} & 
\multicolumn{1}{c}{$\mathcal{E}$} & 
\multicolumn{1}{c}{$F_{VHE}^0$} & 
\multicolumn{1}{c}{$L$} \\

        \hline
          J0308.4+0408  & 0.013 & NGC 1218 &  \textbf{0.03}  &Sy 1  & 0.83&$8.0\times10^{-9}$&1.82&4.4&4.5 &1.8\\
          J0847.0-2337   & 0.12 &  	PMN J0847-2337  & \textbf{0.059} & BLL &1.27 &  $1.3\times10^{-5}$ & 1.65 & 2.7 &7.4 & 11.6\\
          J1037.6+5711 &  0.05 & GB6J1037+5711 &  ?\tnote{1} & BLL & 4.08 &  $1.7\times10^{-6}$ &1.72&5.7&3.5 & ?\\
          J1203.3-3924 &   0.10& 	PMN J1203-3926& 0.28 &BLL  &  0.44 &$5.5\times10^{-5}$ &1.38 & 1.4 & 14.2 &  503\\
          J1457.8-4642 & 0.10  &PMN J1457-4642& 0.16   & BLL? & 0.17 &$1.2\times10^{-4}$ & 1.49 & 1.6 &12.5 &145\\
         ---& 0.007 & NGC 5549?  & \textbf{0.025}  & LIN & ---& $3.3\times10^{-4}$ & 2.16 & 3.2 & 6.2 & 1.7 \\
         \bottomrule
    \end{tabular}
    \begin{tablenotes}
    \item[1] there are several redshift estimation for GB6J1037+5711, we discuss it in more details in the pertinent section below
    
    \end{tablenotes}
    \end{threeparttable}

\end{table}

\twocolumn
\section{Lists of events for transients}

\begin{table}[h!]

\setlength{\tabcolsep}{2pt}

    \centering
    \begin{tabular}{@r^c^r^r^c}
    \toprule
       \multicolumn{1}{c}{$t$, s} & 
\multicolumn{1}{c}{$E$, GeV} & 
\multicolumn{1}{c}{$l$, $^\circ$} & 
\multicolumn{1}{c}{$b$, $^\circ$} & 
\multicolumn{1}{c}{Distance, $^\circ$} \\
    \midrule

         \textbf{560582016}& \textbf{115.9} &  \textbf{174.840}& \textbf{-44.514} & \textbf{0.013}\\
        \textbf{560764691}& \textbf{121.9} &  \textbf{174.841} &\textbf{ -44.511}     & \textbf{0.012}\\
    \bottomrule    
    \end{tabular}
    \caption{List of VHE photons for 3FHL J0308.4+0408 (NGC 1218). Bold rows indicate the doublet events. Columns: (1) Mission elapsed time (seconds); (2) Energy (GeV); (3,4) Galactic coordinates; (5) Angular distance from source center (degrees).}
    \label{tab:0308_vhe}

    \hfill
\end{table}
    \vspace{-1cm}
\begin{table}[h!]
\setlength{\tabcolsep}{4pt}
    \centering
    \begin{tabular}{@r^c^r^r^c}
    \toprule
    \multicolumn{1}{c}{$t$, s} & 
\multicolumn{1}{c}{$E$, GeV} & 
\multicolumn{1}{c}{$l$, $^\circ$} & 
\multicolumn{1}{c}{$b$, $^\circ$} & 
\multicolumn{1}{c}{Distance, $^\circ$} \\
    \midrule
    555018647 &  54.0 &   174.857	&-44.554  &   0.04 \\
    \bottomrule
    \end{tabular}
    \caption{List of 10-100 GeV photons for 3FHL J0308.4+0408 (NGC 1218) in $2\times10^7$~s interval centered on doublet. Columns are the same as in Table \ref{tab:0308_vhe}   Events, temporally coinciding with the doublet, are shown in boldface.}
    \label{tab:0308_10GeV}

\end{table}

\vspace{-0.5cm}

\begin{table}[h!]
\setlength{\tabcolsep}{2pt}
    \centering

    \begin{tabular}{@r^c^r^r^c}
    \toprule
     \multicolumn{1}{c}{$t$, s} & 
\multicolumn{1}{c}{$E$, GeV} & 
\multicolumn{1}{c}{$l$, $^\circ$} & 
\multicolumn{1}{c}{$b$, $^\circ$} & 
\multicolumn{1}{c}{Distance, $^\circ$} \\
    \midrule
257661702    &    491.4 & 247.803  & 12.184 &  0.03 \\
363076249 &      106.8 & 247.665 & 12.145&  0.12 \\ 
370057437 &        135.9  & 247.703 & 12.195 &  0.07 \\
448525521 &        291.9  &  247.926 & 12.149   & 0.16\\
478319482 &        207.6  & 247.738 & 12.270& 0.08 \\
541749219 &        189.0   & 247.771 &  12.254 & 0.05 \\
575995109 &      103.7  &  247.789 & 12.205 &  0.01 \\
\rowstyle{\bfseries}
655392252  &      204.6  &247.615 & 12.252 & 0.16 \\
\rowstyle{\bfseries}
655557713 &         130.2 &  247.731 & 12.268&  0.08 \\
689089028 &      136.0  & 247.911 & 11.979 & 0.26 \\
728372459  &      156.4  &247.815 & 12.217 &0.04 \\
\bottomrule
    \end{tabular}
    \caption{List of VHE photons for 3FHL J0847.0-2337 (PMN J0847-2337), analogous to Table \ref{tab:0308_vhe} }
    \label{tab:0847_vhe}
\end{table}


\begin{table}[H]
\vspace{-10pt}
\setlength{\tabcolsep}{2pt}
    \centering
    \begin{tabular}{@r^c^r^r^c}
    \toprule
    \multicolumn{1}{c}{$t$, s} & 
\multicolumn{1}{c}{$E$, GeV} & 
\multicolumn{1}{c}{$l$, $^\circ$} & 
\multicolumn{1}{c}{$b$, $^\circ$} & 
\multicolumn{1}{c}{Distance, $^\circ$} \\
    \midrule
    648643223 &	20.8 &	247.661 &	12.191	&	 0.11\\
659048631 &88.6& 247.906	&12.286	& 0.15 \\
659088756&	17.1& 247.829	& 11.993 & 0.22 \\
659316050 &	26.9 &	247.787	& 12.157&  0.05 \\
659636820 & 11.0 & 247.635	& 12.318 & 0.18 \\
659881515 & 93.8 & 	247.742 & 	12.269 & 0.07 \\
661263355 & 	14.0 & 	247.816 & 	12.237 &  0.05 \\
662222468 & 	45.2 & 	247.743	 & 12.229 & 	 0.04\\
662302983 &	13.8 & 247.564 & 	12.177&  0.21 \\  
\bottomrule
\end{tabular}
    \caption{List of 10-100 GeV photons for 3FHL J0847.0-2337 (PMN J0847-2337)  analogous to Table \ref{tab:0308_10GeV}.}
    \label{tab:0847_10GeV}
\end{table}

\begin{table}[h!]

\setlength{\tabcolsep}{2pt}
    \centering
    \begin{tabular}{@r^c^r^r^c}
    \toprule
        \multicolumn{1}{c}{$t$, s} & 
\multicolumn{1}{c}{$E$, GeV} & 
\multicolumn{1}{c}{$l$, $^\circ$} & 
\multicolumn{1}{c}{$b$, $^\circ$} & 
\multicolumn{1}{c}{Distance, $^\circ$} \\
    \midrule
    486429674& 126.6 &151.735& 51.806 &0.03\\
511813676 & 133.3 & 151.767 & 51.813 & 0.03\\
515627358 & 135.0 & 151.816 & 51.823 & 0.04\\
518976121 & 118.8 & 151.714 & 51.817 & 0.05 \\
\rowstyle{\bfseries}
551111293 & 126.1 & 151.708 & 51.773 & 0.04 \\
\rowstyle{\bfseries}
551283348 & 169.7 & 151.696 & 51.823 & 0.06\\
584157729& 118.9 & 151.800   & 51.763 & 0.03 \\
650447843 & 104.0 & 151.747& 51.736 &0.05\\
654983499& 133.0 &151.822 &51.822 &0.05\\
735040791& 245.4 &151.888 &51.652 &0.15\\
\bottomrule
    \end{tabular}
    \caption{List of VHE photons for 3FHL  J1037.6+5711 (GB6J1037+5711), analogous to Table \ref{tab:0308_vhe} }
    \label{tab:1037_vhe}
\end{table}

\begin{table}[h!]
\setlength{\tabcolsep}{2pt}
\centering
\begin{tabular}{@r^c^r^r^c}
    \toprule
    \multicolumn{1}{c}{$t$, s} & 
\multicolumn{1}{c}{$E$, GeV} & 
\multicolumn{1}{c}{$l$, $^\circ$} & 
\multicolumn{1}{c}{$b$, $^\circ$} & 
\multicolumn{1}{c}{Distance, $^\circ$} \\
    \midrule
542669658& 23.3 & 151.758 & 51.834 & 0.05\\
546965261& 25.0 & 151.717 & 51.802 & 0.04\\
546998880& 30.4 & 151.662 & 51.851 & 0.10\\
547089708& 51.1 & 151.808 & 51.806  & 0.03\\
547552717& 17.3 & 151.844 & 51.778  & 0.04\\
549223585& 64.3 & 151.809 & 51.799  & 0.03\\
549798866& 11.0 & 151.459 & 51.705  & 0.21\\
550164818& 17.4 & 151.820 & 51.788  & 0.03\\
\rowstyle{\bfseries}
551152678& 11.3 & 151.379 & 51.886  & 0.26\\
551401970& 62.1 & 151.682 & 51.696 &0.10\\
551607468& 10.9 & 151.745 & 51.754 & 0.03\\
553354233& 10.2 & 151.479 & 51.964 & 0.26\\
553993869& 16.3 & 151.726 & 51.796 & 0.03\\
554927761& 27.2 & 151.758 & 51.838 & 0.06\\
555345377& 16.8 & 151.833 & 51.724 & 0.07\\
555517519& 13.7 & 151.679 & 51.767 & 0.06\\
557514043& 12.3 & 151.670 & 51.756 & 0.07\\
558648352& 10.8 & 151.640 & 51.691 & 0.12\\
558744923& 37.3 &152.147 &51.617& 0.29\\ 
\bottomrule
\end{tabular}
    \caption{List of 10-100 GeV photons for 3FHL  J1037.6+5711 (GB6J1037+5711)  analogous to Table \ref{tab:0308_10GeV}.}
    \label{tab:1037_10GeV}
\end{table}

\begin{table}[h!]
\setlength{\tabcolsep}{2pt}
    \centering
    \begin{tabular}{@r^c^r^r^c}
    \toprule
     \multicolumn{1}{c}{$t$, s} & 
\multicolumn{1}{c}{$E$, GeV} & 
\multicolumn{1}{c}{$l$, $^\circ$} & 
\multicolumn{1}{c}{$b$, $^\circ$} & 
\multicolumn{1}{c}{Distance, $^\circ$} \\
    \midrule
    260094483 &  101.2& 293.114 & 22.563 & 0.21 \\
\rowstyle{\bfseries}
373271798 & 172.2&  292.963 & 22.533 & 0.07 \\
\rowstyle{\bfseries}
373409583 & 203.3 & 292.754 & 22.634 & 0.19 \\
694676662 & 101.9& 292.877 & 22.501 & 0.02 \\
\bottomrule
    \end{tabular}
    \caption{List of VHE photons for 3FHL J1203.3-3924 (PMN J1203-3926), analogous to Table \ref{tab:0308_vhe} }
    \label{tab:1203_vhe}
\end{table}

\begin{table}[H]
\setlength{\tabcolsep}{2pt}
    \centering
    \begin{tabular}{@r^c^r^r^c}
    \toprule
     \multicolumn{1}{c}{$t$, s} & 
\multicolumn{1}{c}{$E$, GeV} & 
\multicolumn{1}{c}{$l$, $^\circ$} & 
\multicolumn{1}{c}{$b$, $^\circ$} & 
\multicolumn{1}{c}{Distance, $^\circ$} \\
    \midrule        \rowstyle{\bfseries}
391732356 & 158.7 & 324.526 & 10.815 & 0.15\\
        \rowstyle{\bfseries}
391881102 & 148.2 & 324.357 & 11.010 & 0.16 \\
\bottomrule
    \end{tabular}
    \caption{List of VHE photons for 3FHL J1457.8-4642 (PMN J1457-4642), analogous to Table \ref{tab:0308_vhe} }
    \label{tab:1457_vhe}
\end{table}

\begin{table}[H]
\setlength{\tabcolsep}{4pt}
\centering
\begin{tabular}{@r^c^r^r^c}
    \toprule
    \multicolumn{1}{c}{$t$, s} & 
\multicolumn{1}{c}{$E$, GeV} & 
\multicolumn{1}{c}{$l$, $^\circ$} & 
\multicolumn{1}{c}{$b$, $^\circ$} & 
\multicolumn{1}{c}{Distance, $^\circ$} \\
    \midrule388126965 & 17.7 &  324.369 & 10.825 & 0.02 \\
388309964 & 55.8& 324.365 & 10.840  & 0.01 \\
389170862 & 30.9 & 324.383 & 10.801  & 0.05 \\
389618335 & 10.8 & 324.459 & 10.743  & 0.13 \\
\bottomrule
\end{tabular}
    \caption{List of 10-100 GeV photons for 3FHL J1457.8-4642 (PMN J1457-4642)  analogous to Table \ref{tab:0308_10GeV}.}
    \label{tab:1457_10GeV}
\end{table}

\begin{table}[H]

\setlength{\tabcolsep}{2pt}
    \centering
    \begin{tabular}{@r^c^r^r^c}
    \toprule
     \multicolumn{1}{c}{$t$, s} & 
\multicolumn{1}{c}{$E$, GeV} & 
\multicolumn{1}{c}{$l$, $^\circ$} & 
\multicolumn{1}{c}{$b$, $^\circ$} & 
\multicolumn{1}{c}{Distance, $^\circ$} \\
    \midrule
    \rowstyle{\bfseries}
580411826& 156.0 & 353.234 & 61.364 & 0.04 \\
        \rowstyle{\bfseries}
580628550 &  592.9 &  353.25 &  61.437 & 0.03\\
\bottomrule
    \end{tabular}
    \caption{List of VHE photons for NGC5549(?), analogous to Table \ref{tab:0308_vhe} }
    \label{tab:ngc5549_vhe}
\end{table}

\end{document}